%% file: main.tex
\documentclass[a4paper,11pt]{article}
\usepackage{a4wide}

\usepackage{amsmath}
\usepackage{amssymb}
\usepackage{placeins}
\usepackage{graphicx}
\usepackage{tabularx}
\usepackage{tabulary}
\usepackage{multirow}
\usepackage{authblk}

\usepackage[
  natbib=false,
  sorting=none, 
  bibstyle=numeric-comp,
  citestyle=numeric-comp,
  backend=bibtex,
  defernumbers
]{biblatex}

\DeclareFieldFormat{url}{\url{#1}}

\renewbibmacro{in:}{}

\DeclareFieldFormat[article]{pages}{#1}

\DeclareFieldFormat*{title}{``\textit{#1}''}
\DeclareFieldFormat{journaltitle}{#1}

\DeclareFieldFormat{doi}{%
\newline
\mkbibacro{DOI}\addcolon\space
\ifhyperref
{\href{http://dx.doi.org/#1}{\nolinkurl{#1}}}
{\nolinkurl{#1}}}

\renewbibmacro{in:}{%
\ifentrytype{article}{}{%
\printtext{\bibstring{in}\space}}}
\DeclareBibliographyDriver{set}{\entryset{%
\ifnumequal{\thefield{entrysetcount}}{1}{}{\addsemicolon\space}}{}%
\finentry}

\usepackage{hyperref}

\title{Flavour-changing neutral scalar interactions of the top quark}
\author[1]{N. F. Castro}
\author[2]{K. Skovpen}

\affil[1]{Laborat\'{o}rio de Instrumenta\c{c}\~{a}o e F\'{i}sica
Experimental de Part\'{i}culas (LIP),
Departamento de F\'{i}sica, Escola de Ci\^{e}ncias,
Universidade do Minho, 4710-057 Braga, Portugal}
\affil[2]{Ghent University, Sint-Pietersnieuwstraat 33, 9000 Gent,
Belgium}

\date{}

\begin{document}

\maketitle

\begin{abstract}
A study of the top quark interactions via flavour-changing
neutral current (FCNC) processes provides an intriguing connection between
the heaviest elementary particle of the standard model (SM) of particle
physics and the new scalar bosons that are predicted in several
notable SM extensions. The production cross sections of the processes with
top-scalar FCNC interactions can be significantly enhanced to the
observable level at the CERN Large Hadron Collider. The present 
review summarizes the latest experimental results
on the study of the top quark interactions with the Higgs boson via
FCNC and describes several promising directions to look for new
scalar particles.
\end{abstract}

\flushbottom
\newpage

\input{introduction}
\input{experiment}
\input{bsm}
\input{future}
\input{summary}

\section*{Acknowledgements}

NC is supported by FCT - Funda\c{c}\~{a}o para a
Ci\^{e}ncias e a Tecnologia, Portugal, under project
CERN/FIS-PAR/0032/2021.

\newpage
\printbibliography

\end{document}

%% file: introduction.tex
\section{Introduction}

Conservation laws and flavour-symmetry structures represent the core
element of any theoretical model that provides a description of
interactions involving elementary particles. An experimental study
of fundamental interactions is an excellent probe of higher-order 
symmetries, potentially leading to a construction of a more complete
model of nature with resolving the remaining unanswered questions of the remarkably successful
Standard Model (SM). The SM theory of weak interactions allows
flavour-violating processes in the quark sector through the charged
weak currents. Such flavour-changing transitions proceed via an exchange 
of a W boson between the two fermionic states. The weak eigenstates are 
treated as left-handed doublets, allowing transitions between the up- and down-type quarks, while the mass
eigenstates are represented by a superposition of weak eigenstates
connected via an unitary matrix. The rotation from one type of state to
another is then expressed as the Cabibbo-Kobayashi-Maskawa (CKM)
matrix which governs the flavour-mixing processes through the flavour-changing charged weak
transitions~\cite{KomMasWeak}. The processes, where a fermion changes 
its flavour via an exchange of a neutral boson, are therefore absent at
tree level in the SM due to the unitarity of rotational matrices and are called flavour-changing neutral currents
(FCNC)~\cite{GlaIliWeak}.

The effect of flavour mixing in the quark sector was first introduced using a three-quark model that only included the
$\mathrm{u}$, $\mathrm{d}$, and $\mathrm{s}$ quarks~\cite{CabUni}.
Experimental studies of the $\mathrm{K_{L} \to \mu^{+}\mu^{-}}$
decays and neutral kaon mixing processes however indicated important
difficulties in satisfying theoretical predictions for FCNC transitions~\cite{MaiGim}.
The problem was theoretically solved in 1970s by introducing the
fourth type of quark, the charm ($\mathrm{c}$) quark, in order to restore the
quark-lepton symmetry of the weak interaction. It was shown that an
additional contribution associated with an exchange of a $\mathrm{c}$ quark at
one-loop level almost completely cancels the respective contributions
connected to the lighter quarks. This effect leads to a significant suppression of
FCNC transitions at higher orders --- the Glashow-Iliopoupos-Maiani (GIM) mechanism.
The discovery of the c quark, just a few years later, confirmed these theoretical 
speculations~\cite{AugRes, AubJps}. The four-quark model was later
extended to include five quark flavours, after the discovery of the bottom ($\mathrm{b}$) quark~\cite{HerDim}. It
took a bit longer for the top ($\mathrm{t}$) quark to be experimentally
observed in 1995, completing the SM to contain six quark
flavours~\cite{AbeTop, AbaTop}. In a full representation of the quark
sector, the tree-level
transitions between different quark flavours are only allowed through
the weak flavour-changing charged interaction, while the tree-level FCNC transitions are
completely missing in the SM and are only possible as loop corrections.

The FCNC effects are predicted in the leptonic sector as lepton flavour
violating transitions. However, the probability to observe such
processes is expected at the level of $\simeq~10^{-54}$, in the case of the $\mu \to
e\gamma$ decay, due to an extreme suppression from the neutrino mass
difference to the power of four and is experimentally inaccessible~\cite{GouFla, CelFla, CalFla}.
The FCNC transitions in the decays of hadronic states with
$\mathrm{s}$, $\mathrm{c}$, or $\mathrm{b}$ quarks are observed experimentally~\cite{AubPho, NikEta,
MohGam, LinKao, LeeCha, BatKao}. The studies of these processes are however
affected by the large uncertainties in theoretical calculations of branching 
ratios of hadron decays, mainly driven by the
non-perturbative long-distance strong interaction contributions. 

The lifetime of the top quark ($\mathrm{\tau_{t} \simeq 5 \times
10^{-25}\ s}$), that is shorter than the typical
formation time of the bound states ($\mathrm{\tau_{had} =
1/\Lambda_{QCD} \simeq 10^{-24}\ s}$), makes the processes with the top quark
production an excellent probe to search for FCNC effects.
The absence of hadronic activity leading to
the formation of bound states involving top quarks makes the study of
the FCNC processes less affected by radiative QCD corrections. 
The FCNC effects can be probed in the top quark production 
processes, as well as in the decays of the top quarks.
The amplitude of an FCNC transition is proportional to the squared
mass of the quark involved in the loop diagram. A remarkable suppression 
of the top quark FCNC decays is explained by the fact that the only possible one-loop contributions 
are associated with the lighter quarks, leading to the branching fractions of 
$\mathrm{\mathcal{B}(t \to cX) \simeq 10^{-15} - 10^{-12}}$~\cite{AguTop}, 
where $\mathrm{X}$ represents either a gluon ($\mathrm{g}$), photon
($\mathrm{\gamma}$), $\mathrm{Z}$ or a Higgs boson ($\mathrm{h}$). Theoretical predictions
for the top quark FCNC effects are available with the next-to-leading 
order (NLO) precision~\cite{ZhaNlo, ZhaTop}, as well as the approximate
next-to-next-to-next-to-leading order calculations for some of these
processes~\cite{ForTop, GuzTop}.

The study of the flavour structure of the SM is one of the strongest
probes of the beyond the SM (BSM) theories. A strong suppression of the top
quark FCNC transitions is a perfect condition to search for various
possible deviations from the SM predictions. Several experimental studies of the properties of
FCNC decays of b hadrons have sparked a series of intriguing anomalies
in the measured probabilities of the rare $\mathrm{b \to s
\ell^{+}\ell^{-}}$ FCNC transitions, as well as in the measurements of
the ratios $\mathrm{\mathcal{B}(B^{+} \to K^{+}\mu^{+}\mu^{-})/\mathcal{B}(B^{+} \to
K^{+}e^{+}e^{-})}$~\cite{AaiBme},
$\mathrm{\mathcal{B}(B^{0} \to K^{*0}\mu^{+}\mu^{-})/\mathcal{B}(B^{0} \to
K^{*0}e^{+}e^{-})}$~\cite{AaiUni}, as well as the branching
fractions~\cite{AaiIso, AaiWav, AaiAng}.
A common analysis of these results reveals a potential tension with 
respect to the SM~\cite{AmiAno, CapPat, ArbHad, KimAno, BanAno}.
Experimental searches for FCNC effects in the top quark sector
represent therefore an important channel to probe the 
anomalous interactions of the third-generation quarks.

%% file: experiment.tex
\section{Experimental studies of the top quark FCNC processes}
\label{sec:exp}

The top quark FCNC effects can be probed directly in the production of
a single top quark, as well as in the top quark decays.
Studies of the top quark FCNC decays are typically associated with 
similar sensitivities to the top quark FCNC couplings with an up and a charm quark. 
Experimental sensitivities to these couplings
mainly differ in terms of the performance of various reconstruction methods
used for identification of hadronic jets originating from quarks of
different flavour. At hadron colliders, the single top quark FCNC production process is
mostly sensitive to the top quark FCNC coupling with an up quark (or an up
antiquark) due to an enhanced sensitivity due to the proton distribution function of
the colliding protons (or antiprotons).
The importance of these two production channels depends 
on a specific type of the top FCNC coupling that is probed in an experiment.

Before the LHC, the top FCNC couplings were studied in electron-positron collisions at
LEP2~\cite{ArcSin, AbbSin, BarSin, AbdSin},
in deep inelastic scattering processes at HERA~\cite{AktGlu, CheGlu, AbrSin,
AarSin, HeiAle}, and in
proton-antiproton collisions at Tevatron~\cite{AbaGlu, AbaGlu2,
AalZbo, AalGlu}. 
The electron-positron colliders allow for a study of the top-$\mathrm{\gamma}$ and top-Z couplings
in the processes with the production of a single top quark,
$\mathrm{e^{+}e^{-} \to t\bar{c}(\bar{u})}$. The study of deep inelastic scattering
of electrons on protons has an enhanced sensitivity to
the same type of couplings in the processes of $\mathrm{ep \to et+X}$, as well
as to the top-gluon FCNC couplings in the $\mathrm{ep \to etq(g)+X}$ processes.
The obtained experimental constraints were recently improved
by almost one order of magnitude after the analysis of the LHC proton-proton 
collision data~\cite{SirZbo, KhaGlu, KhaPho, ChaFcn, ChaZbo,
AabZbo, AadGlu, AadZbo, AadZbo7, AadPho}. 

The top-Higgs FCNC transitions receive the largest suppression in the
SM with respect to the other top quark FCNC processes because of the large 
mass of the Higgs boson. These transitions
are among the rarest processes predicted in the SM in the quark
sector, and therefore, the study of these processes is associated with
a generally enhanced sensitivity to potential new physics effects. The discovery of the Higgs boson at
the LHC paved a way to a comprehensive study of the top-Higgs FCNC processes
at the ATLAS and CMS experiments, which resulted in the
first experimental constraints on these anomalous 
couplings~\cite{AabHig, AabHigML, AabHigGG, AadHig8, KhaHig8, SirHig, CMSHGGR2, CMSHBBR2, ATLASHTTR2}.
The direct searches for the top-Higgs FCNC effects are performed in top quark
decays, as well as in the associated production of single top quarks with
a Higgs boson. Many of the performed studies were targeting the top quark 
FCNC decays in $\mathrm{t\bar{t}}$ events. In recent studies of the 13~TeV data, 
the analysis of the single top quark associated production with a Higgs boson was also
included~\cite{SirHig, CMSHGGR2, CMSHBBR2, ATLASHTTR2}.

\subsection{$\mathrm{h \to \gamma\gamma}$}

Search channels that are relevant to the top-Higgs
FCNC couplings are usually defined based on the Higgs boson decay channels.
The Higgs boson decays to pairs of photons provide a clean experimental environment to look
for the top-Higgs FCNC effects. In addition to the two
photons, these final states consist of up to one isolated lepton with
additional hadronic jets. The analysis strategy is primarily based on the
reconstruction of the Higgs boson di-photonic invariant mass.
The contributions from various background processes are fitted in the
mass sidebands in data, followed by its 
extrapolation to the signal region. In these fits, the background
contributions that are associated with the
SM Higgs boson production must be accounted for, representing one of the
dominant resonant backgrounds in the search region.
The uncertainty associated with the choice of the fit function,
the statistical uncertainty in data, as well as the background contributions from the SM processes
involving the Higgs boson, represent the main uncertainties in the
study of these final states.

The searches for top-Higgs FCNC processes in the $\mathrm{h \to
\gamma\gamma}$ channel were carried out by ATLAS~\cite{AabHigGG} and
CMS~\cite{CMSHGGR2} in the single-lepton
and hadronic final states, including a pair of photons. The
integrated luminosity of recorded 13 TeV data corresponds to
36~fb$^{-1}$ and 137~fb$^{-1}$, respectively.
The identification of isolated photon objects and the common
vertex of the photon pair is the core part of the
analysis. The photon and the common vertex identification algorithms 
are based on the multivariate analysis (MVA) approaches. The
obtained mass resolution allows to observe a resonance
structure in the diphoton invariant mass spectra in simulated signal
events corresponding to the Higgs boson decay. The contributing
nonresonant background processes include the diphoton production with jets, as
well as the top quark pair and the vector boson production processes
with additional photons. The SM production of the Higgs boson
represents the dominant resonant background. The nonresonant
backgrounds are estimated directly from data by performing a fit to
the reconstructed diphoton invariant mass spectrum. The fitted function represents the sum of a
double-sided Crystal Ball function that corresponds to the signal
prediction, the resonant background from the SM Higgs production, and a parameterized function describing the nonresonant
background obtained in a data control region. The main uncertainties include the b tagging and jet
energy corrections, as well as photon identification systematic
uncertainties. The uncertainty in the limited number of
events in data also represents an important limiting factor in the
final sensitivity in these searches.
An additional contribution to the total systematic
uncertainty is associated with theoretical uncertainties in
the prediction of the resonant background processes with the SM Higgs
boson production. The unbinned likelihood fit to data using the described
signal and background diphoton mass spectra is performed, and the constraints are
set on the top quark FCNC decay branching fractions.
The observed (expected) limits obtained by ATLAS are $\mathcal{B}(\mathrm{t \to hc}) <$~2.2~$\times$~10$^{-3}$
(1.6~$\times$~10$^{-3}$) and $\mathcal{B}(\mathrm{t \to hu})
<$~2.4~$\times$~10$^{-3}$ (1.7~$\times$~10$^{-3}$). The observed
(expected) constraints obtained in the CMS analysis are 
$\mathcal{B}(\mathrm{t \to hc}) <$~7.3~$\times$~10$^{-4}$
(5.1~$\times$~10$^{-4}$) and $\mathcal{B}(\mathrm{t \to hu})
<$~1.9~$\times$~10$^{-4}$ (3.1~$\times$~10$^{-4}$). An enhanced sensitivity
obtained in the CMS analysis is explained by a larger data
sample used in the study, as well as due to the inclusion of the
top-Higgs FCNC process with an associated production of a single top quark
and a Higgs boson. The latter has led to an improved sensitivity to
$\mathcal{B}(\mathrm{t \to hu})$.

\begin{figure}[!htbp]
\centering
\includegraphics[width=.46\textwidth]{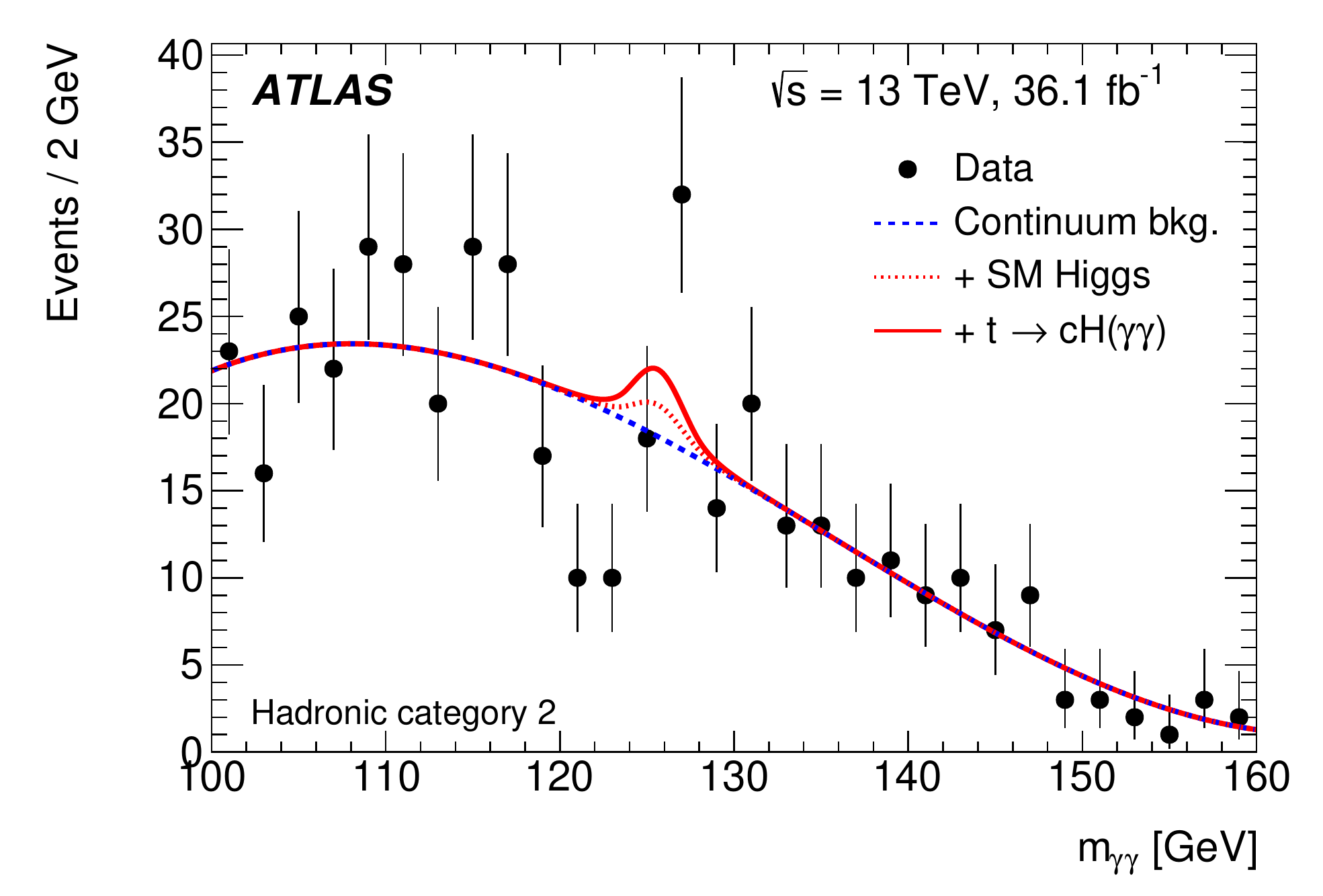}
\includegraphics[width=.46\textwidth]{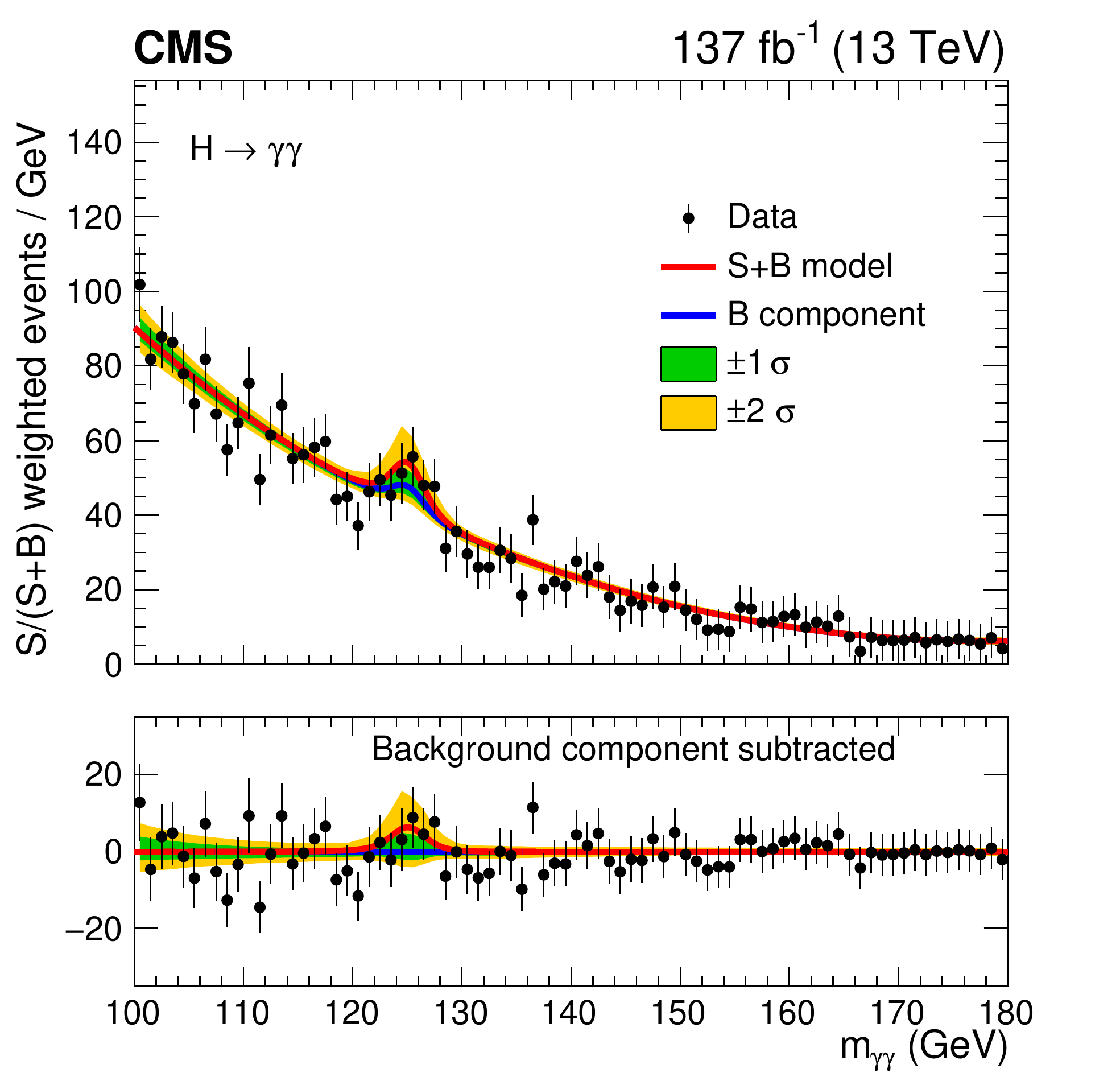}
\caption{\label{fig:ExpGG}
Distributions with the invariant diphoton mass showing the results of the fit to
data in the top-Higgs FCNC study of the $\mathrm{h \to \gamma\gamma}$
channel at (left) ATLAS~\cite{AabHigGG} and (right)
CMS~\cite{CMSHGGR2}. The ATLAS results are presented for hadronic final states, while the
CMS results include a combination of all considered channels
with events weighted by the associated significance of each event
category.}
\end{figure}

\subsection{$\mathrm{h \to WW/ZZ/\tau\tau}$}

Multilepton final states arise from the
Higgs boson decays to a pair of W or Z bosons, as well as to $\tau$ leptons.
Event categories in these studies are
associated with the final states including two same-sign and three
leptons. The same-sign lepton channel has the dominant background
contributions originating from the processes with nonprompt and
misidentified leptons, while the three-lepton channel is mainly affected by
the presence of diboson events as well as nonprompt leptons.
These backgrounds are estimated from data. The search channels involving one hadronic
$\tau$ lepton identified in the Higgs boson decay receive dominant
background contributions from the processes with misidentified
$\tau$ lepton decays, as well as from events with the SM production
of top quarks. In the case when the decays of both $\tau$ leptons result in
hadronic final states, a significant background contribution is also
associated with the Z boson decays to the pairs of $\tau$ leptons.

\begin{figure}[!htbp]
\centering
\includegraphics[width=.46\textwidth]{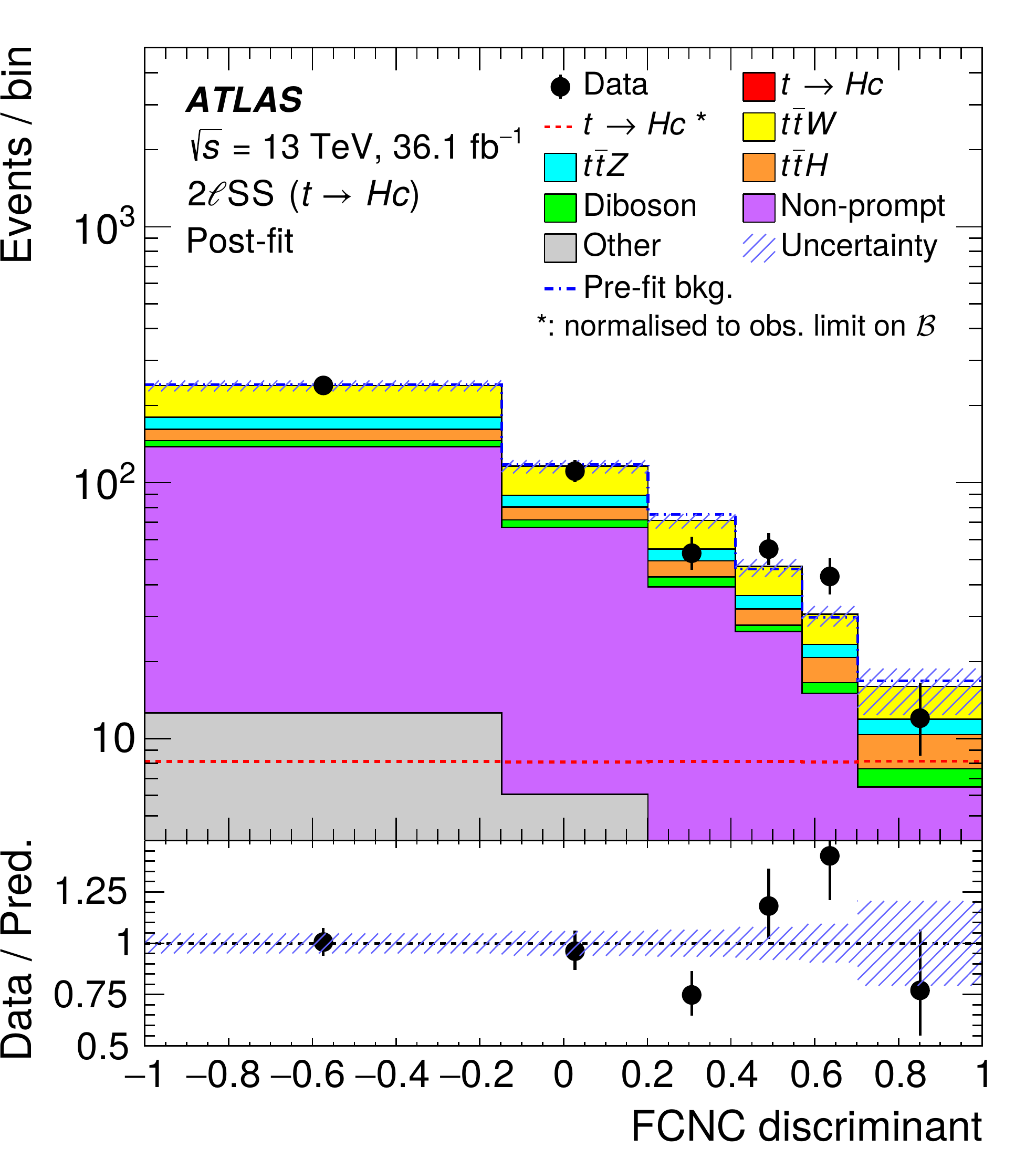}
\includegraphics[width=.46\textwidth]{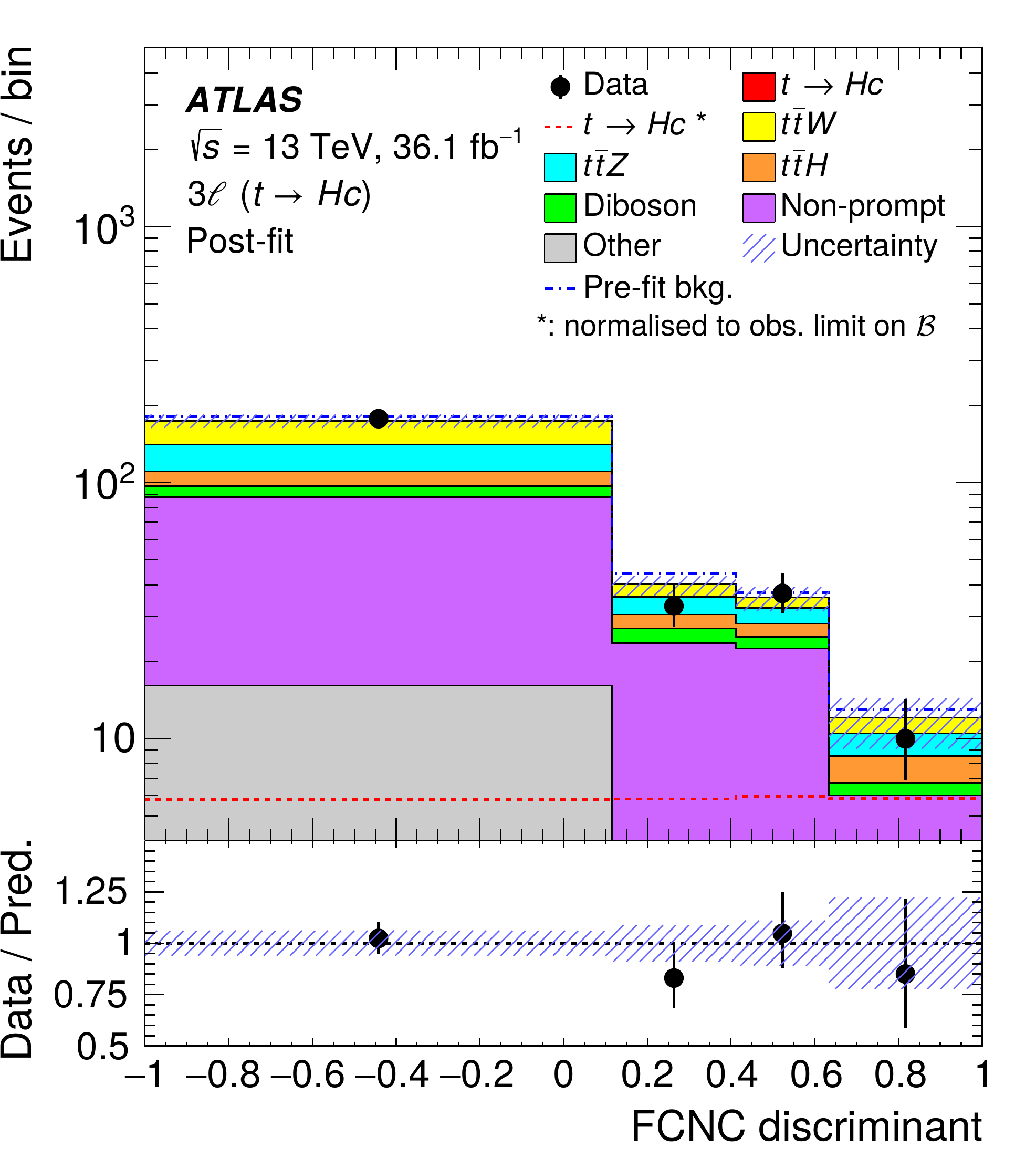}
\caption{\label{fig:ExpML}
Distributions of the BDT discriminants using the top-Higgs FCNC signal
selection in multilepton search channels in the same-sign dilepton (left) and trilepton (right) final
states~\cite{AabHigML}. The presented BDT discriminant was optimized
for the case of the $\mathrm{t \to hc}$ FCNC decays.}
\end{figure}

The searches for top-Higgs FCNC couplings in the multilepton channels
were performed at ATLAS~\cite{AabHigML} and CMS~\cite{KhaHig8} using
36~fb$^{-1}$ of 13 TeV and 20~fb$^{-1}$ of 8 TeV data, respectively.
Events are split into the final states with two same-sign (2lSS)
and three (3l) leptons. The dominant backgrounds are associated with
the nonprompt and misidentified leptons, as well as with the leptons
originating from photon conversions. The prompt-lepton backgrounds
correspond to events with an associated production of top quark pairs
and a W, a Z, or a Higgs boson, with additional contributions arising from the processes
with diboson production. The baseline selection criteria require the 
presence of two or three leptons and at least two jets, with one or
two b-tagged jets. The prompt lepton
identification plays an important role in these studies in suppressing
the dominant nonprompt lepton backgrounds. The statistical and
systematic uncertainties associated with the prediction of the backgrounds
with nonprompt leptons are among of the dominant uncertainties in
these searches.
Two separate boosted decision tree (BDT) discriminants involving various reconstructed
kinematic variables are trained in the 2lSS and 3l channels to further
suppress various background. The BDT distributions that are presented in Fig.~\ref{fig:ExpML} are
used in a binned maximum-likelihood fit to extract the constraints on the top-Higgs FCNC processes.
The observed (expected) 95\% CL
limits on the top quark FCNC branching fractions in the multilepton
final states 
$\mathcal{B}(\mathrm{t \to hc}) <$~1.6~$\times$~10$^{-3}$
(1.5~$\times$~10$^{-3}$) and $\mathcal{B}(\mathrm{t \to hu})
<$~1.9~$\times$~10$^{-3}$ (1.5~$\times$~10$^{-3}$) are obtained.
Multilepton searches provide an excellent sensitivity to the top-Higgs FCNC
couplings, however, the existing results use only a partial data set, 
and further updates on these studies are anticipated in the future.

A dedicated study of the top-Higgs FCNC effects in final states with
one or two hadronically decaying $\tau$ leptons was recently done by
ATLAS with 139~fb$^{-1}$ of 13 TeV data~\cite{ATLASHTTR2}. The
analysis strategy is similar to the one used in the previous
analysis~\cite{AabHig}, with an increased number of kinematic regions
sensitive to the signal production, in order to account for the
single-top production channel for top-Higgs FCNC. The dominant
background in this search is associated with the presence of nonprompt
$\tau$ leptons, estimated from data. Other backgrounds are predicted
by simulation. The obtained constraints on the branching fractions are
$\mathcal{B}(\mathrm{t \to hc}) <$~9.4~$\times$~10$^{-4}$
(4.8~$\times$~10$^{-4}$) and $\mathcal{B}(\mathrm{t \to hu})
<$~6.9~$\times$~10$^{-4}$ (3.5~$\times$~10$^{-4}$).

\subsection{$\mathrm{h \to b\bar{b}}$}

The Higgs boson decays to a pair of b
quarks with the largest branching fraction of $\simeq$~58\%~\cite{LHCXS1}.
A considerable amount of background events is associated with the $\mathrm{t\bar{t}}$
production with additional hadronic jets. The analysis of
this channel is systematically limited with the dominant contributions
to the total uncertainty arising from the
application of the heavy flavour jet identification techniques, as
well as the modelling uncertainties relevant to the predictions of the
top quark production processes with additional jets. One of the
important handles in suppressing background processes is the kinematic
event reconstruction involving top quarks and additional jets.
The assignment of reconstructed final-state objects to the initial hard-process 
particles is done using the MVA methods.

\begin{figure}[!htbp]
\centering
\includegraphics[width=.46\textwidth]{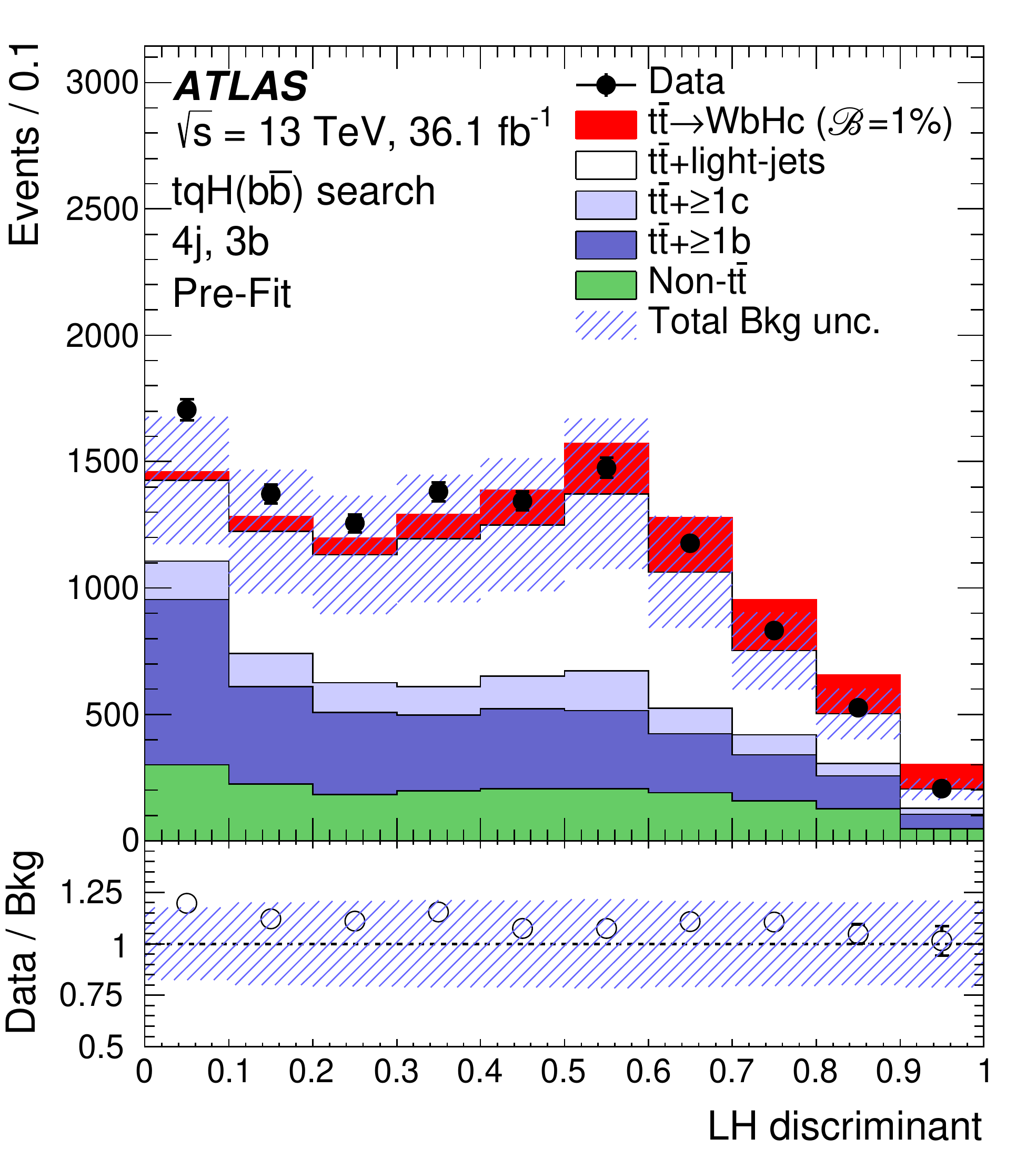}
\includegraphics[width=.46\textwidth]{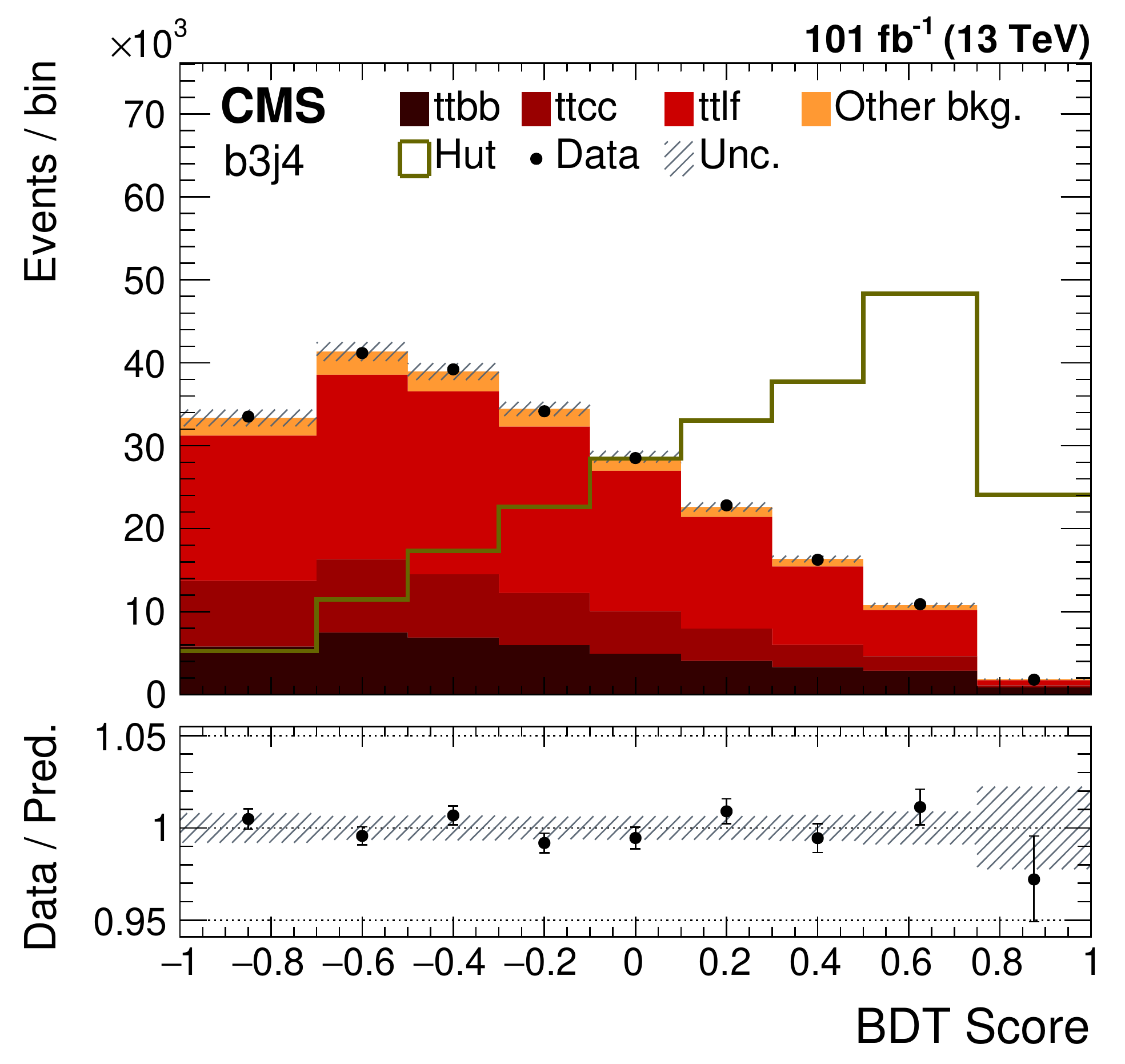}
\caption{\label{fig:ExpBB} Distributions of the discriminants used in
the analyses of the $\mathrm{h \to
b\bar{b}}$ channel at (left) ATLAS~\cite{AabHig} and (right)
CMS~\cite{CMSHBBR2}. Selected events correspond to the final states with
four reconstructed jets, three of which are identified as associated
with heavy-flavour hadron decays. The pre-fit and post-fit results are
shown for ATLAS and CMS, respectively.}
\end{figure}

The top FCNC search in the $\mathrm{h \to b\bar{b}}$ channel is
performed in the final states with one isolated lepton and additional
jets. The total integrated luminosity used in the ATLAS
analysis corresponds to
36~fb$^{-1}$ of 13 TeV data~\cite{AabHig}. The CMS results use
101~fb$^{-1}$ of data~\cite{CMSHBBR2},
additionally combined with the previously published result from the
analysis of 36~fb$^{-1}$ data~\cite{SirHig}.
The ATLAS analysis focuses on the study of the event topology
with at least four jets in the final state, mainly relevant to the top
quark FCNC decays. The corresponding CMS analysis additionally includes the 
signal top quark production mode of signal events, and therefore,
the requirement on the minimum number of reconstructed jets is set to
a lower value.
At least three b-tagged jets are required to be present in event. In
both analyses the selected events are classified based on the number of
jets and b-tagged jets. The dominant background contributions correspond to the top
quark pair production in association with light-flavour jets in the
event categories with two b-tagged jets, while the associated
production of top quark pairs with heavy-flavour jets
($\mathrm{t\bar{t}b\bar{b}}$, $\mathrm{t\bar{t}c\bar{c}}$) represents the dominant
background in the case of the higher number of b-tagged jets. Theoretical predictions 
for these processes are subject to relatively large uncertainties due
to the renormalization and the factorization scale variations arising from
the different energy scales of the top quark mass and the jet
transverse momentum involved in the generation process, 
as well as the inclusion of heavy quark masses in the calculations~\cite{ThttHF1}.
The experimental uncertainties in the measurement
of the production cross sections of these processes reach $\simeq$~10--20\%~\cite{AtlttHF1, CmsttHF1, CmsttHF2, CmsttHF3}.
The background processes are further suppressed by using the
discriminants that exploit kinematic information of the selected
reconstructed objects, defining the probability of an event
to correspond to the signal process hypothesis. As shown in
Fig.~\ref{fig:ExpBB}, in the ATLAS analysis this is done by constructing the likelihood (LH) discriminant, while
the BDT approach is used in the case of the CMS search. The
binned maximum-likelihood fits are performed to data based on the described
discriminants to extract the limits on the FCNC
contributions, resulting in the observed (expected) 95\% CL
constraints on the top quark FCNC branching fractions of
$\mathcal{B}(\mathrm{t \to hc}) <$~4.2~$\times$~10$^{-3}$
(4.0~$\times$~10$^{-3}$) and $\mathcal{B}(\mathrm{t \to hu})
<$~5.2~$\times$~10$^{-3}$ (4.9~$\times$~10$^{-3}$). The resultant constraints
in the CMS analysis are $\mathcal{B}(\mathrm{t \to hc})
<$~9.4~$\times$~10$^{-4}$ (8.6~$\times$~10$^{-4}$) and
$\mathcal{B}(\mathrm{t \to hu}) <$~7.9~$\times$~10$^{-4}$
(1.1~$\times$~10$^{-3}$). The differences in the sensitivities in the published results
by the two experiments are mainly due to the different size of the
analyzed data sample, as well as to the inclusion of the single top quark
production mode for the top-Higgs FCNC process in the case of the CMS analysis. 
A combination of the results obtained from the
analyses of different Higgs boson decay channels, $\mathrm{h \to
\gamma\gamma}$, $\mathrm{h \to WW/ZZ}$, $\mathrm{h \to
\tau\tau}$ and $\mathrm{h \to b\bar{b}}$, was performed at
ATLAS using 36~fb$^{-1}$ of data, corresponding to the limits of 
$\mathcal{B}(\mathrm{t \to hc}) <$~1.1~$\times$~10$^{-3}$
(8.3~$\times$~10$^{-4}$) and $\mathcal{B}(\mathrm{t \to hu})
<$~1.1~$\times$~10$^{-3}$ (8.3~$\times$~10$^{-4}$)~\cite{AabHig}.
The ATLAS constraints on the top-Higgs FCNC interactions are competitive to the ones obtained in
the analysis of the $\mathrm{h \to \gamma\gamma}$ channel at
CMS~\cite{CMSHGGR2}, which already uses all available recorded
data at 13 TeV.

\subsection{Indirect searches}

The top-Higgs FCNC interactions can be indirectly constrained from the
studies of the SM processes that can potentially include FCNC
loop-level contributions involving top quarks.
The relevant processes include the hadron electric
dipole moments~\cite{HarHig, GorHig}, $\mathrm{Z \to c\bar{c}}$
decays~\cite{LarFcn} and  $\mathrm{D^0-\overline{D^0}}$
mixing~\cite{FerHig}. The indirect limits are competitive with the current
direct constraints obtained at the LHC~\cite{HesHig}.

%% file: bsm.tex
\section{Global approach to the FCNC searches}

A broad range of experimental searches for new physics phenomena have been using the
$\kappa$-framework to parameterize the potential deviations from
the SM predictions~\cite{KapFra}. This framework defines a set of scaling factors
for production cross sections and decay widths as a function of the
new physics model parameters. While the $\kappa$-framework has proved to be
very successful in theoretical interpretation of a large number of
experimental results, it does not represent an ultimate approach to provide a 
complete systematic description of various new physics effects.

Given the absence of any strong evidence of new physics, the natural
assumption is that new particles are much heavier than the SM
particles, and its direct production at present is not achievable
within the LHC energy range. The potentially induced new
physics effects at the electroweak scale can be parameterized with a
general effective field theory (EFT) approach that includes additional
high energy dimension operators in the extended Lagrangian of the SM
(SMEFT). The rich phenomenology of the SMEFT includes 59 independent operators,
assuming baryon number conservation~\cite{BucEff, GrzEff}. A full categorization of
EFT operators relevant to the top quark sector and its interplay with
other SM processes is summarized in Refs.~\cite{AguEff, AguEff2, AguEff3}. 
Several of these operators are relevant to the FCNC processes with top quarks. 
The Wilson coefficients (WCs) of the respective EFT operators can be constrained from
the measured production cross sections, as well as from the study of
the shapes of various kinematic variables. An EFT analysis of experimental 
observables represents a general approach to study potential deviations 
from the SM predictions that can be used to set constraints on various BSM models.
The top quark FCNC EFT couplings comprise several dimension-6
operators, which are discussed in Ref.~\cite{AguEff2, AguEff3, DegEff, DurEff}.
The FCNC EFT contributions can also interfere at higher orders~\cite{DegEff, DurEff, CasTop}.

Potential FCNC EFT effects with top quarks were probed in experimental studies
of top-gluon and top-photon FCNC processes at the LHC~\cite{EftDil, AadPho}.
Additionally, several of the obtained experimental constraints
on the FCNC top quark decay branching fractions and strength
$\kappa$-modifiers, which were described in Section~\ref{sec:exp},
can be directly translated into the corresponding limits on the
relevant WCs. The first direct measurement of the constraints on the EFT WCs relevant to
the top-Higgs FCNC interactions was recently performed in Ref.~\cite{ATLASHTTR2}.
Re-interpretations of various
experimental results that are sensitive to the top-Higgs EFT operators 
are also available~\cite{DurEff, AguEff}.

\section{New scalar bosons}

Several extensions of the SM can induce sizable FCNC effects that can
be experimentally probed at the LHC. There are two possible ways to
introduce top-Higgs FCNC in a BSM model. The first possibility is to
increase the number of fermions, modifying the CKM matrix structure, 
escaping the GIM suppression. This approach is usually referred to
as the minimal flavour violation (MFV). The second option is to
involve new heavy particles in the loops of the higher order diagrams,
increasing the probability of FCNC transitions. The study of top-Higgs
FCNC effects appears to be rather promising in various simplest extensions of
the SM, where the additional neutral scalar particles can potentially
mix with the SM Higgs boson.    

A dedicated estimate of the BSM-enhanced branching fractions of $\mathrm{t
\to u(c)h}$ decays shows the maximal values reaching $10^{-3}-10^{-4}$
in some of the BSM models~\cite{AguTop, BarDet}. Such high event rates are
being probed at the LHC using recorded data with the typical constraints 
set at the level of $\simeq 10^{-3}$. However,
the maximal branching fractions are not necessarily associated with the
most favorable parameter space of a BSM model and can potentially
involve additional model-tuning. 

The SM Higgs boson can have its additional partners in various BSM
scenarios. The Two-Higgs Doublet Model (2HDM) is one of the simplest
extensions of the SM that introduces two Higgs doublets with five
scalar particles: $\mathrm{h^{0}}$, CP-odd $\mathrm{A^{0}}$, CP-even
$\mathrm{H^{0}}$ ($\mathrm{m_{H} > m_{h}}$),
and $\mathrm{H^{\pm}}$, where $\mathrm{h^{0}}$ is the lightest CP-even SM-like
Higgs boson~\cite{BraDbl, GunDbl, MaiDbl}. The 2HDM contains seven parameters, with only two
of them relevant at leading order (LO), usually defined as
$\mathrm{cos(\beta-\alpha)}$ and $\mathrm{tan(\beta)}$. The former
parameter is related to the couplings of a scalar particle to vector
bosons, while the latter represents the ratio of the vacuum 
expectation values of the heavy and the SM Higgs bosons.

There are four (I, II, III, and IV) types of 2HDM. The 2HDM-I and
2HDM-II do not include FCNC processes at tree level due to the
requirement of flavour conservation via the presence of a
$Z_{2}$-symmetry. In these two types of 2HDM all fermions couple to
the same Higgs boson. In the 2HDM-III without an imposed discrete
symmetry the fermions can couple to both Higgs doublets, and the
tree-level FCNC transitions involving the top-charm FCNC couplings
with the Higgs boson can be significantly enhanced~\cite{HouFcn}. 
A combined fit of various results from the direct and indirect 
experimental searches favors the alignment limit
$\mathrm{cos(\beta-\alpha) \to 0}$ with $\mathrm{cos(\beta-\alpha) <
0.1-0.4}$, with some additional dependence on $\mathrm{tan(\beta)}$, the mass of the scalar
boson, as well as the type of the model~\cite{SirCom, AabCom, AabHea, PicDbl,
AbbDbl, DasDbl, KaoTop, CheTop, JaiTop, BauTop, BraCkm, BotMin, BotDbl}. 
The alignment scenario corresponds to the case, when $\mathrm{h^{0}}$ and 
the SM Higgs boson share the same couplings.

In the aligned two-Higgs-doublet model (A2HDM) it is assumed that both
Yukawa matrices are aligned in flavour space to avoid FCNC at
tree level~\cite{PicDbl}. The enhanced one-loop-induced $\mathrm{t \to
ch}$ decays can occur in such models~\cite{AbbDbl}. 
Special extensions of the 2HDM models can incorporate FCNC at tree level, 
such as the top quark 2HDM (T2HDM)~\cite{DasDbl}.
In this model it is assumed that the top quark is the only elementary
fermion that couples to the non-SM Higgs doublet to generate its large
mass in a natural way, therefore allowing the top-Higgs FCNC due to a small
$\mathrm{cos(\beta-\alpha)}$ admixture of the exotic neutral Higgs boson. The
study of the $\mathrm{t \to c h}$ decays represents a promising channel to
probe the T2HDM and 2HDM-III at the LHC~\cite{KaoTop, CheTop, JaiTop, BauTop}. 
There are also the so-called BGL modifications of the 2HDM,
where the tree-level top-Higgs FCNC transitions can be associated with
either up- or down-type quarks, preserving the structure of the
CKM matrix~\cite{BraCkm, BotMin, BotDbl}.

Additional Higgs doublets can naturally appear in the context of
supersymmetric (SUSY) theories. The Minimal Supersymmetric Standard
Model (MSSM) is the simplest extension of the SM representing the
2HDM-II with an additional supersymmetric particle
content~\cite{DjoMss, FaySup, FaySup2}.
At tree level, this model contains two non-SM parameters: the mass of
the CP-odd Higgs boson, $\mathrm{m_{A}}$, and $\mathrm{tan(\beta)}$. 
An effective MSSM model with the lightest CP-even SM Higgs boson is referred to as
hMSSM, where the properties of the SM Higgs boson define the remaining
masses and couplings of the MSSM~\cite{MaiDbl, DjoMss2, DjoMss3}. 
This approximation of MSSM is only completely valid at the moderate
values of $\mathrm{tan(\beta)}$. The recent LHC experimental searches 
generally disfavor small values of $\mathrm{m_{A}}$ below $\simeq$ 600 GeV 
within the hMSSM~\cite{SirCom, AabCom, AabHea, PicDbl, AbbDbl, DasDbl, 
KaoTop, CheTop, JaiTop, BauTop, BraCkm, BotMin, BotDbl}. The predicted
top-Higgs FCNC rates in a general MSSM can reach $10^{-7}$~\cite{DedMss},
while in case of the R-parity violation (RPV) in a general SUSY model
these transitions can be enhanced to $10^{-5}$~\cite{EilSus}.

An extended MSSM with baryon (B) and lepton (L) numbers as local
symmetries, broken near the electroweak scale, is known as
BLMSSM~\cite{PerBar, PerBar2, PerBar3, AmoBar}. This model can
incorporate an enhancement of $\mathrm{t \to ch}$
rates at one-loop~\cite{GaoTop}. The Next-to-Minimal Supersymmetric Standard
Model (NMSSM) represents an extension of MSSM that naturally
generates the mass parameter $\mu$ in the Higgs superpotential at the
electroweak scale and resolves the so-called
$\mu$-problem~\cite{EllNms, KimMup}.
The new neutral scalars considered in the MSSM theories can
potentially mix with the SM Higgs boson and therefore generate top-Higgs
FCNC at tree level. 

The addition of exotic vector-like quark to the CKM matrix allows to
escape the GIM mechanism. The top-Higgs FCNC transitions can be
enhanced to $10^{-5}$ in the Quark Singlet (QS)~\cite{AguExo} and Alternative
Left-Right Models (ALRM)~\cite{GaiAlt}. Similar enhancements can be achieved
in the Littlest Higgs Model with T-parity (LHT) induced by
interactions with the new T-odd gauge bosons and fermions~\cite{YanTop}. The
presence of Kaluza-Klein fermion states in the Randall-Sundrum (RS)
models with warped extra dimensions can produce sizable FCNC
effects of the same order~\cite{AzaDim, CasRsm}.

The new light neutral scalar singlets (S) are present in various
supersymmetric extensions of the SM, including NMSSM and the Composite
Higgs Models (CHM)~\cite{DimMas, KapMis, KapCom, CacSca, CasNov}, with
$\mathrm{t \to cS}$ tree-level FCNC decays~\cite{BanExt}. In such extensions these scalars are considered as
Nambu-Goldstone bosons (pNGBs) with the Higgs boson, representing a
bound state of a new strongly-interacting dynamics. The large mass of
the top quark can be generated through the mixing of elementary fermions
with a composite operator of a high scaling dimension~\cite{KapFla}. In CHM,
the SM elementary particles can be seen as composite states that mix
with its heavy partners. This model provides a promising explanation
of the mass hierarchy of the SM by introducing a new physics scale and
the idea of compositeness of the SM particles. The $\mathrm{t \to cS}$ decays, with
$\mathrm{m_{S} < m_{t} - m_{c}}$, are expected to strongly dominate over
the $\mathrm{t \to ch}$ transitions in CHM, providing a new window to constrain
the new physics models via top quark FCNC searches with neutral light scalars. 
These processes are not yet studied experimentally. The
predicted rates of the $\mathrm{t \to cS}$ decays can be probed down to
$10^{-5}$ with the existing LHC data~\cite{CacSca, CasNov}.

In addition to the top quark FCNC decays with a Higgs boson, one can also search 
for FCNC decays of the heavy neutral scalars (H) predicted in many BSM models. 
In a general 2HDM model, as well as in its extensions, such as T2HDM, the
probability of the $\mathrm{H \to t\bar{c}}$ decay is proportional to
$\mathrm{sin(\beta-\alpha)}$, while the probabilities of the
$\mathrm{t \to ch}$ decays are proportional to
$\mathrm{cos(\beta-\alpha)}$~\cite{AltFla}. This represents an
important complementarity of the top FCNC searches in the top quark and
heavy neutral scalar decays. At high energies, one of the dominant
decays of the heavy scalars is the production of two top
quarks, if the mass of the scalar particle exceeds the doubled mass of
the top quark. However, in the heavy scalar mass range of 175 and 350
GeV, the $\mathrm{H \to t\bar{c}}$ decays are associated with the 
largest branching fraction in the model parameter space, favored by the current experimental
constraints~\cite{GorHea}. A study of the $\mathrm{H \to t\bar{c}}$
decays is a promising way to search for heavy neutral scalar particles at the LHC.
The dominant production mode for heavy scalars is expected to be the
gluon-gluon fusion process, however, in the context of the
``flavourful'' 2HDM (F2HDM), that removes the 2HDM-intrinsic $Z_{2}$
discrete symmetry and additionally modifies the structure of Yukawa
matrices~\cite{KnaHie, AltLoc, AltCol}, the dominant channel is the single top associated
production with a heavy neutral scalar ($\mathrm{pp \to tH \to tt\bar{c}}$),
resulting in the presence of two same-sign top quarks in the final
state~\cite{AltRar}. The searches for the heavy scalar FCNC decays
with top quarks are also proposed within the Froggatt-Nelsen Singlet
Model (FNSM), mostly relevant for the HL-LHC data analysis~\cite{ArrFla}.

%% file: future.tex
\section{Future perspectives}

The LHC has accumulated about 25 fb$^{-1}$ of proton-proton collision
data at 7 and 8 TeV, as well as nearly 140 fb$^{-1}$ of data at 13
TeV. The latest studies from the LHC on the top quark FCNC processes
therefore focus on the analysis of the 13 TeV data.
The next round of the data taking at the LHC is planned for 2022,
where it is expected that the total accumulated statistics will be
doubled, reaching 300 fb$^{-1}$ by the end of the LHC project.
The future experiments at the High-Luminosity LHC (HL-LHC) are
planned to bring almost an order of magnitude larger
data set of 3 ab$^{-1}$ due to a significant increase in the
instantaneous luminosity of colliding proton beams up to $10^{35}$
cm$^{-2}$ s$^{-1}$, representing a 5 to 7 times higher luminosity
with respect to its nominal value. The projected
sensitivities of the top quark FCNC searches, following the preliminary
estimates of the expected performances of the upgraded ATLAS and CMS detectors
at the HL-LHC, indicate a significant improvement in the constraints on
the branching fractions of the top quark FCNC decays after the analysis of
the full LHC statistics reaching an order of magnitude~\cite{AbaFCC}.

There is a number of major international projects under consideration
in the high-energy physics domain, defining an evolving strategy for 
this field for many years to come. The Large Hadron
electron Collider (LHeC) is proposed as an extension of the LHC
project, re-using the existing proton accelerator complex and combining it with a
new electron accelerator for production of 60 GeV electron beams for
the study of the deep inelastic scattering at high
energies~\cite{AbeLhe, BruLhe, KleLhe, KleDis}.
The planned experiments at the LHeC are mostly sensitive to the
top-$\mathrm{\gamma/Z}$ FCNC couplings, and the projected limits are 
expected to be comparable to the corresponding sensitivities at
the HL-LHC~\cite{CakLhe, BehLhe, CakLhe2}. The study of the top-Higgs FCNC
interactions appears to be less promising at the LHeC, the expected sensitivity of
which has been already surpassed by the latest LHC results~\cite{WanLhe}.

The electron-proton collisions are considered as part of the
Future Circular Collider (FCC) project, involving several experiments
targeting different types of high-energy collisions. The FCC-eh
machine will collide 60 GeV electron with 50 TeV proton beams,
produced by the FCC accelerator~\cite{BorLhe, BruJow}. Due to the increased energy of
the proton beams, relatively similar sensitivities for the top-Higgs
FCNC couplings are expected to the ones of the HL-LHC.
Most of the improvements are anticipated for the top-$\mathrm{\gamma/Z}$ FCNC
couplings~\cite{AbaFCC, BehFcc}. The planned experiments at the high precision electron-positron
collider (FCC-ee) will also be very sensitive to the
top-$\mathrm{\gamma/Z}$ FCNC couplings~\cite{AbaFCC2}. 
The dominant sensitivity to the top-Higgs FCNC processes at the FCC-ee
is mainly associated with the top quark decay channels. 
The FCC-hh machine with proton-proton collisions at $\sqrt{s}$ = 100 TeV will 
allow increased sensitivity to all relevant top quark FCNC couplings,
probing the $\mathrm{t \to ch}$ decay branching fractions down to
$\simeq 10^{-5}$~\cite{AbaFCC3, ManFCC, PapFcc, OyuFcc, OyuFcc2}.
The High Energy LHC (HE-LHC) project will adopt the FCC-hh technology to use 
proton-proton collisions at $\sqrt{s}$ = 27 TeV. The HE-LHC is viewed as capable 
to improve the HL-LHC limits on the top quark FCNC couplings by an order of
magnitude~\cite{AbaFCC4, LiuHel}. The linear electron-position
colliders, such as ILC and CLIC, are also associated with good
prospects for the top quark FCNC studies~\cite{MooPol, ZarCli, ZarCli2, AbrCli, MelHig}.
However the projected ILC/CLIC sensitivities for the top-Higgs FCNC interactions
are not expected to reach the sensitivity level of the corresponding
studies at the HL-LHC.

\begin{figure}[!htbp]
\centering
\includegraphics[width=.80\textwidth]{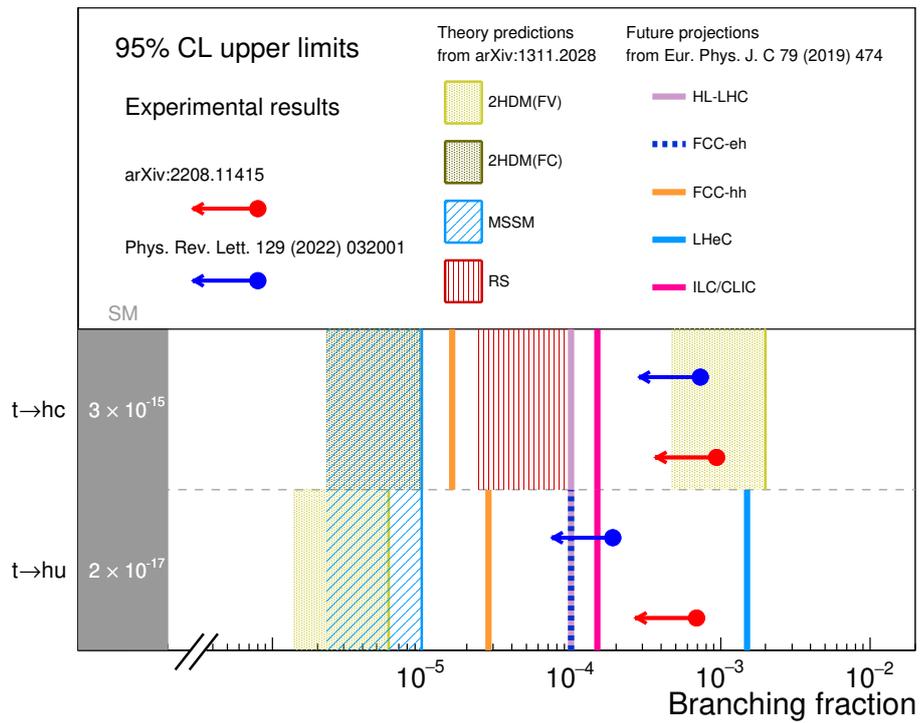}
\caption{\label{fig:fcncProj} Summary of the best experimental
constraints to date on the top-Higgs FCNC processes at the
LHC~\cite{ATLASHTTR2, CMSHGGR2},
including sensitivity projections for
future colliders. The results are also compared to various BSM
predictions that correspond to the maximal expected branching
fractions in a given model. Adapted from Ref.~\cite{AbaFCC}.}
\end{figure}

A summary of the described experimental results and future
projections is presented in Fig.~\ref{fig:fcncProj}. The analysis of
the $\simeq$~140~fb$^{-1}$ of the LHC 13 TeV data allows to reach the 95\% confidence
level limits of the order of 10$^{-4}$ and 10$^{-3}$ for the $\mathrm{t
\to uh}$ and $\mathrm{t \to ch}$ decay branching fractions, respectively.
These experimental limits are obtained from the analysis of the
the Higgs boson decays to photon pairs, and therefore, the presented 
results are expected to be further improved when
combined with the results obtained in the analysis of other
Higgs boson's decay modes. The illustrated sensitivities for future
colliders are also obtained in the analyses of specific channels. The
comparison with the ultimate sensitivities is expected to be more complete, 
once the LHC and the future projection results become available for
all relevant decay channels of the Higgs boson. Based on the considered 
projections, the best expected sensitivity of $\simeq$~10$^{-5}$ is associated with 
the experiments at the FCC-hh.

The described sensitivities of the future experiments mainly correspond to the
studies of the top-Higgs FCNC couplings, with only a few projections
available for some of the top quark flavour-changing neutral scalar
processes. While the searches for new scalars via top quark FCNC
appear to be highly relevant for the HL-LHC, as well as its successors, these
processes are not yet explored at full extent with the existing LHC data.

%% file: summary.tex
\section{Summary}

The sensitivity of the LHC experiments has been defying the predictions 
of various BSM models for the maximal top quark FCNC couplings.
The study of the top quark FCNC processes involving the neutral scalar bosons is an
excellent probe of the new physics effects in a number of most promising BSM scenarios, 
including additional Higgs doublets and 
scalar singlets through partial compositeness. In some models the production of new
scalars can be significantly enhanced by the flavour-changing neutral
scalar couplings, and therefore, these studies represent a very 
promising direction to look for additional heavy and light partners of the 
discovered Higgs boson. Beyond the LHC, the upcoming experiments at
the HL-LHC and FCC are expected to come with even better sensitivities to probe
the top quark anomalous couplings with the new scalars. The analysis
of the LHC and the future collider data will remain the only way to
directly probe the top quark flavour-changing neutral scalar interactions 
in the next decades.